\documentclass[12pt]{iopart}

\usepackage{graphicx}

\def\be{\begin{equation}}
\def\ee{\end{equation}}
\def\bea{\begin{eqnarray}}
\def\eea{\end{eqnarray}}

\def\EF{$E_{\rm F} $}
\def\kF{$k_{\rm F} $}
\def\hv{h$\nu $}
\def\kx{k$_x$}
\def\ky{k$_y$}

\def\a{$\alpha $}

\def\A-1{\AA$^{-1}$}
\def\LPB{Li$_{0.9}$Mo$_6$O$_{17}$}
\def\~{$\approx$}
\def\hv{$h\nu $}

\begin{document}

\title[PES and the Unusually Robust 1d Physics of \LPB]{Photoemission Spectroscopy and the \newline Unusually Robust One Dimensional Physics of \newline Lithium Purple Bronze}

\author{L. Dudy$^{1,2}$, J. D. Denlinger$^3$, J. W. Allen$^{1}$, F. Wang$^{1,4}$, J. He$^{5}$, D. Hitchcock$^{5}$, A. Sekiyama$^{6}$, S. Suga$^{6}$}

\address{$^1$ Randall Laboratory, University of Michigan, Ann Arbor, MI 48109, USA}

\address{$^2$ Physikalisches Institut, Universit\"at W\"urzburg, D-97074 W\"urzburg, Germany}

\address{$^3$ Advanced Light Source, Lawrence Berkeley National Laboratory, Berkeley, CA, 94270, USA}

\address{$^4$ BTG Pactual, 601 Lexington Avenue -- 57th floor, New York, NY 10022 USA}

\address{$^6$ Department of Physics and Astronomy, Clemson University, Clemson, SC 29634, USA}

\address{$^6$ Department of Material Physics, Graduate School of Engineering Science, Osaka University, 1-3 Machikaneyama, Toyonaka, Osaka 560-8531, Japan}

\date{\today}

\begin{abstract}
Temperature dependent photoemission spectroscopy in
Li$_{0.9}$Mo$_6$O$_{17}$ contributes to evidence for one dimensional
physics that is unusually robust.  Three generic characteristics of
the Luttinger liquid are observed, power law behavior of the
k-integrated spectral function down to temperatures just above the
superconducting transition, k-resolved lineshapes that show holon
and spinon features, and quantum critical (QC) scaling in the
lineshapes.  Departures of the lineshapes and the scaling from
expectations in the Tomonaga Luttinger model can be partially
described by a phenomenological momentum broadening that is
presented and discussed.  The possibility that some form of 1d
physics obtains even down to the superconducting transition
temperature is assessed.

\end{abstract}

\maketitle

\section{Introduction}

\LPB, the so-called lithium purple bronze (hereafter called LiPB),
has emerged as a paradigm quasi one-dimensional (1d) material for
studying the stability of Luttinger liquid chains against single
particle electron hopping $t_{\perp}$ between the chains
\cite{viewpoint}.  The usual theoretical understanding \cite{Bois,
Giamarchi} and empirical experience is that quasi-1d materials can
display such 1d behavior at sufficiently elevated temperatures but
that, with decreasing temperature (T), $t_{\perp}$  leads to a
crossover to some kind of 3d behavior, e.g., a Fermi liquid, or an
ordered state such as a charge density wave (CDW) or spin density
wave (SDW).  Because of the excellent Fermi surface (FS) nesting
that is typical of quasi-1d materials, density waves are especially
likely.   Focusing on photoemission spectroscopy but considering
other data as well, this paper explores the possibility that LiPB,
in spite of excellent FS nesting, does not develop a density wave
and crosses over to 3d behavior only when it becomes a
superconductor (SC).  If such is true then the SC could well be
unconventional, as suggested in a recent paper \cite{HusseyHc2}
reporting that the superconducting critical field is far above the
Pauli limit.

LiPB is a material with overall 3d bonding but whose valence band
electronic structure is derived from two parallel zig-zag Mo-O
chains per unit cell.  The chains lie in well separated planes and
receive electrons donated by Li ions located out of the planes.
Early tight binding \cite{Whangbo88} and more recent LDA
\cite{POPOVIC06} band calculations agree that the Mo-O orbitals of
the chains give rise to four bands, two of which (A, B) are always
below the Fermi energy (\EF) and two of which (C,D) cross \EF.  The
resulting FS is quasi-1d but the LDA calculations also show some
splitting and warping  due to single particle hopping $t_{\perp}$
between chains both within a unit cell and in neighboring unit
cells.  Angle resolved photoemission spectroscopy (ARPES) observes
all four bands with the general behavior that is predicted in LDA
calculations \cite{JDD99, Gweon01, WANG06}, including the quasi-1d
FS \cite{JDD99}.  Figures presented in later discussions of ARPES data show the four bands (Fig. \ref{Fig2}(a) in Section 3), and the FS and Brillouin zone (Fig.~\ref{Fig4}(c) in Section 4).  The magnitude of $t_{\perp}$ is discussed in Section 6 of the paper.

The physical properties of LiPB were first investigated in the
1980's \cite{Greenblatt84, Schlenker85, MatsudaXiNoCDWLoc86,
MatsudaMixedCrys86, SatoLoc87, EkinoTunneling87, Schlenker88,
Degiorgi88, SchlenkerPressure89, Schlenker89, Dumas93}.  Electrical
transport \cite{Greenblatt84} was found to be highly anisotropic and
superconductivity (SC) \cite{Greenblatt84, Schlenker85,
EkinoTunneling87, Schlenker88, SchlenkerPressure89} was observed
below T$_{SC}$ = 1.9 K, albeit not in all samples \cite{Schlenker89}
and with a variation of T$_{SC}$ \cite{Greenblatt84, Schlenker85,
EkinoTunneling87, Schlenker88, Schlenker89} that might involve the
Li stoichiometry \cite{He} or disorder \cite{EkinoTunneling87}. More
recent studies \cite{Choi04, Neumeier07, NeumeierThermal07,
Neumeier08, Hussey09, Chen2010} confirm the anisotropic transport
but with differences in the absolute values of the resistivities
along various axes, and also in the magnitudes of the anisotropy
ratios.  These differences may reflect variation in samples and in
the method \cite{Neumeier07} of the measurement.

\section{Resistivity Upturn and Evidence Against a Density Wave}

\begin{figure}[t]
\begin{indented}
\item[]\includegraphics[width=0.65\textwidth] {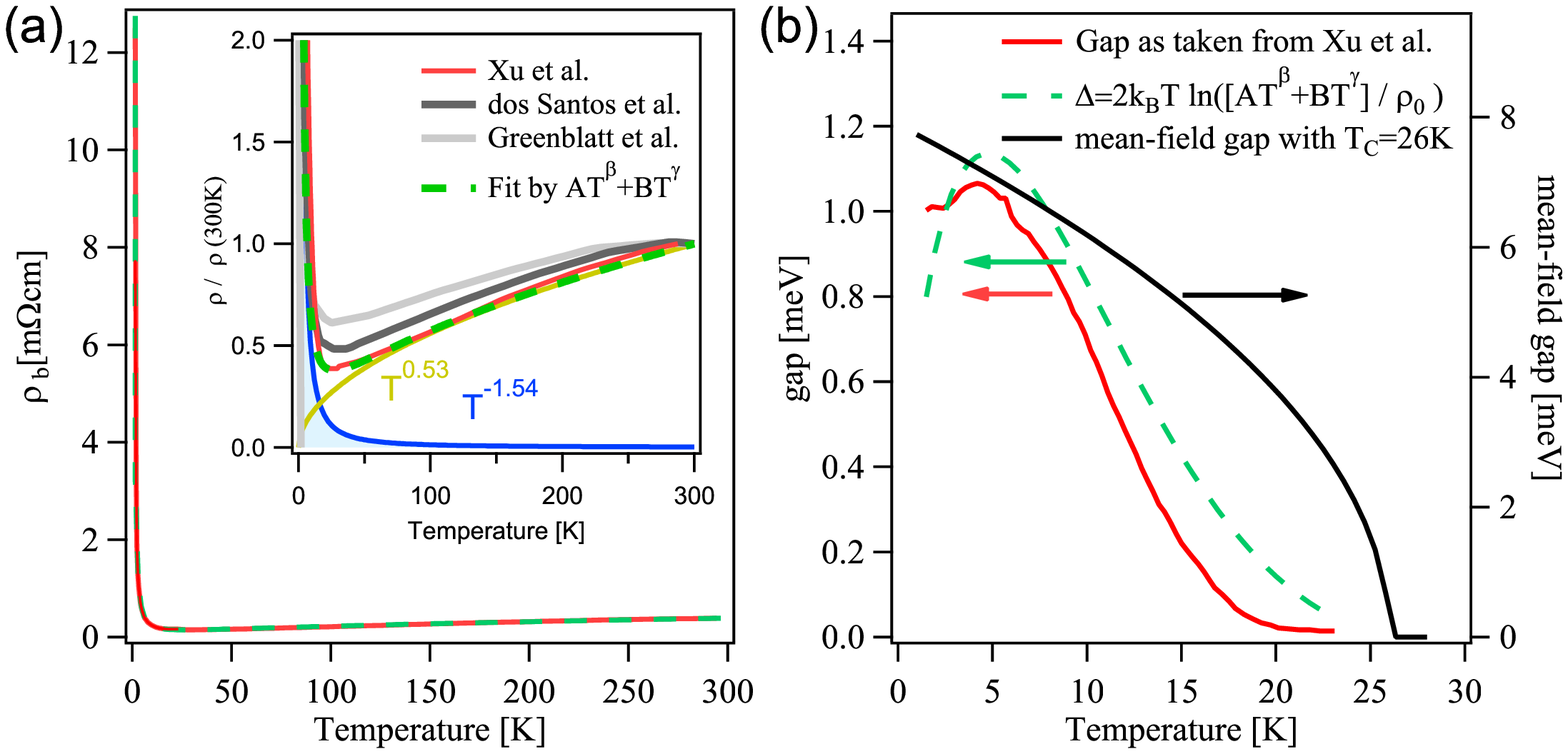}
\caption{Panel (a) shows the temperature-dependent resistivity as taken from the publication of Xu et al. \cite{Hussey09} together with a line-fit of the data with the function $\rho_{Fit}=AT^\beta+BT^\gamma$. The coefficients found were $\beta=0.53\pm0.1$ and $\gamma=-1.54\pm0.2$. The inset shows that the data from Ref.~\cite{Greenblatt84} and Ref.~\cite{Neumeier08} are qualitatively similar.  Note that the inset shows the resistivity normalized to its 300K value in order to be comparable with that of Ref.~\cite{Neumeier08}. Panel (b) shows the gap as extracted in Ref.~\cite{Hussey09} together with the gap extracted here by using our fit $\Delta_{Fit}=2\,k_B T \,ln (\rho_{Fit}/\rho_0)$. Also included is a mean-field temperature behavior of a gap $\Delta_{mf}=3.52\, k_B T_C \, \sqrt{1-T/T_C}$ calculated with $T_C$=26\,K. Note that for the much larger mean-field gap, the axis is on the right.}
\label{Fig1}
\end{indented}
\end{figure}

The early experimental work found that the resistivity is metallic
with decreasing T only down to a minimum at T$_{min}$ $\approx$ 26
K, below which there is an abrupt upturn.  This finding has been
robust over time.  Fig.~\ref{Fig1}(a) shows early data
\cite{Greenblatt84} and more recent data \cite{Neumeier08, Hussey09}
for the upturn as measured along the 1d direction.  An upturn at the
same temperature is also seen for the two directions perpendicular
to the chains.  There is yet no consensus as to the origin of this
upturn because each proposed explanation requires that one or
another piece of data be ignored, be given less weight in the
argument, or be left unexplained \cite{same}.  Early discussions
ascribed the upturn to Anderson localization due to disorder
\cite{MatsudaXiNoCDWLoc86, MatsudaMixedCrys86, SatoLoc87,
EkinoTunneling87} or to a single particle gap, most likely due to
CDW formation \cite{Greenblatt84, Schlenker85, SchlenkerPressure89,
Schlenker89, Dumas93}.  Both direct and nuanced lines of argument
\cite{Greenblatt84, Schlenker85, MatsudaXiNoCDWLoc86, SatoLoc87,
MatsudaMixedCrys86, SchlenkerPressure89, Schlenker89, Dumas93,
Choi04, Neumeier07, NeumeierThermal07, Neumeier08, Hussey09,
Chen2010} for and against these two hypotheses have been presented
many times in the transport literature.  Ref.~\cite{Choi04} is the
most comprehensive recent overview of the issue.

The disorder hypothesis was proposed \cite{MatsudaXiNoCDWLoc86,
SatoLoc87} because a number of important expected signatures of a
density wave are missing in LiPB.  X-ray diffraction studies that
have revealed CDW ordering in related materials \cite{FouryPouget}
have not observed a CDW in LiPB in spite of repeated attempts over
time \cite{Pouget}.  Low-T optical spectroscopy that has observed
single particle gaps in various CDW and SDW materials
\cite{Degiorgi88, TravWach84, Degiorgi96} finds \cite{Degiorgi88,
Choi04} in LiPB no gap even down to 6 K and 1\,meV.  This finding
would seem to be conclusive for the magnitude of gap ($\approx$~8
meV) expected in a mean field description of a CDW transition at
T$_{min}$, as seen from the BCS curve of Fig.~\ref{Fig1}(b).
Comparison to mean field is done here for lack of a better
alternative.  On the one hand, the actual transition temperature for
a quasi-1d system is likely controlled by a small $t_{\perp}$ and so
can be much less than the mean field temperature for a single chain
\cite{Lee73}.  In this case the actual gap could be much larger than
that obtained in this estimate.  On the other hand the single chain
gap for T=0 can be much reduced from its mean field value by LL
fluctuations on the chain, in which case the gap could be smaller
than in this estimate.  In fact Fig.~\ref{Fig1}(b) shows that the
T-dependent gap that is deduced \cite{Hussey09} by describing the
upturn as an Arrhenius law is slightly smaller than 1\,meV at low T
and so could have gone undetected in the optical work.  Such a small
gap was invoked as an explanation for the recent and very exciting
finding \cite{Hussey09} that a sufficiently large magnetic field
applied along the 1d direction can entirely suppress the upturn. The
explanation put forth is that the magnetic field quenches the CDW
and hence the gap because Zeeman splitting of the bands disrupts FS
nesting.  The discussion in Ref.~\cite{Hussey09} also took up a
previous proposal \cite{NeumeierThermal07} for a purely electronic
CDW, driven entirely by Coulomb interactions in a strong coupling
limit.  This proposal was made to overcome indications from optical
spectroscopy \cite{Choi04} and thermal expansion
\cite{NeumeierThermal07} measurements that the lattice has no strong
involvement with the upturn, which implies that a conventional CDW
involving a periodic lattice distortion does not occur.  While such
a purely electronic CDW would be more difficult to detect in the
conventional x-ray diffraction performed for LiPB to date
\cite{Pouget}, it is certainly not impossible as shown by such an
observation for a Bechgard salt \cite{PougetBechgard}.

The great beauty of the experimental magnetoresistance result
\cite{Hussey09} and the neatness of the explanation notwithstanding,
there is no CDW theory that would explain such an unusual gap
function and one must also ignore another important missing
signature of a single particle gap \cite{Greenblatt84,
MatsudaXiNoCDWLoc86, Choi04}, that the dc magnetic susceptibility
measured by three groups over time is entirely unaffected by the
resistivity upturn at T$_{min}$.  Ref. ~\cite{Choi04} points out
that because the magnitude of the measured susceptibility involves a
near cancelation of several large negative and positive
contributions the measurement is particularly sensitive to a change
in any one of them, e.g that of the conduction electrons.  This is
perhaps the strongest single piece of all the many evidences that
collectively weigh heavily against a CDW.  The lack of CDW
signatures in LiPB is all the more striking because the closely
related compounds KMo$_6$O$_{17} $ and NaMo$_6$O$_{17}$ do exhibit
CDW behavior clearly seen with x-ray diffraction \cite{FouryPouget}
and optics \cite{Degiorgi88}.  These materials are also planar but
have higher symmetry such that the planes contain three Mo-O chains
oriented at 120 degrees to one another.  The CDW Q-vector is such as
to gap out the quasi-1d bands associated with two of the three
chains.  Finally the possibility of an SDW in LiPB has been excluded
by muon spin relaxation measurements \cite{muon05}.  The last
section of the paper returns to the issue of the upturn and to the
other parts of Fig.~\ref{Fig1}.

\section{Signatures of Luttinger liquid behavior}

Early workers did not include the ideas of the Luttinger liquid (LL)
in their thinking.  LL behavior in LiPB was first demonstrated by
photoemission experiments.  Indeed, these experiments were
undertaken specifically because of the missing signatures for CDW
formation.  Pioneering studies \cite{Dardel} had observed in angle
integrated photoemission of certain quasi-1d materials one of the
characteristic features expected of LL behavior, that the angle
integrated single particle spectrum approaches \EF\ as a power law.
These studies suffered from some ambiguity because the materials
studied all manifested low T CDW formation and CDW fluctuations at
higher T can in principle cause pseudogap behavior \cite{Lee73,
McKenzie96} that might mimic the expected LL behavior.  Thus LiPB
was conceived \cite{JDD99} as a non-CDW quasi-1d paradigm for
studying LL behavior. It was also envisioned that ARPES experiments
could distinguish features of the single particle spectral function
that are unique to LL behavior. A series of ARPES studies dating
from 1999 \cite{JDD99, Gweon01, WANG06, Gweon02, Allen02, Gweon03, Gweon04,
WANG06PRB, Wang09} has born out this motivation. This
section gives a guide to the key results of these various past ARPES
studies and points out two other measurements using different techniques that have
added very importantly to the case for LL behavior.  Sections 4 and 5 present
new photoemission data and a new phenomenological analysis of ARPES data, respectively.

ARPES measures the single particle spectral function.  For the
paradigm one band Tomonaga-Luttinger (TL) model \cite{TLM} the
spectral function \cite{Orgad01} shows three characteristic
signatures of LL behavior.  The ARPES measurements described below
have demonstrated all three of these signatures.  First, as already
mentioned, the momentum(k)-summed spectrum approaches \EF\ as a
power law characterized by an anomalous exponent \a.  Second, the
k-resolved single particle lineshape is composed of two features, a
holon peak with a leading spinon edge, the two dispersing with
different velocities, $v_c$, $v_s$, respectively.  This lineshape
signifies the absence of quasi-particles and the fractionalization
of the electron into density fluctuation modes of charge (holon) and
spin (spinon).  Third, the TL-model is quantum critical (QC). QC
systems display scale invariance, i.e. at T=0 K and for large
distances and long times, the correlation functions lack a
characteristic scale and take the simplest possible scale-free,
functional form, a power law.  Departures from T=0 K satisfy simple
scaling laws \cite{Hertz76,SACHDEV} with T the only scale.  For the
case of a spin rotationally invariant interaction the TL-model
spectral function \cite{Orgad01} explicitly obeys the ideal scaling
form $A(k,\omega,T) = T^{\eta} \tilde{A}(vk/T, \omega/T)$ where
$\tilde{A}$ is a universal scaling function, $k$ is measured from
the Fermi momentum $k_{\rm F}$, $\omega$ is measured from the Fermi
energy $E_{\rm F}$, and $v$ is a constant with units of velocity.
Within the one band TL-model the T-scaling exponent $\eta$ is then
also determined by a scaling relation to be $\eta = (\alpha-1)$.

The initial ARPES studies \cite{JDD99}  were performed at relatively
high temperatures of 200\,K to 300\,K to avoid any influence of the
putative CDW transition at $T_{min}$ but the TL-model theoretical
lineshapes then available \cite{Meden1992} for comparison
\cite{JDD99,Gweon02} to the ARPES data were for T=0 K. Over time
the quality of the spectra improved and nonzero-T theory lineshapes
\cite{Orgad01} became available.  The latter were particularly
important because including the effect of temperature changed the
parameters deduced from the T=0 K comparison such as to improve the
internal consistency of the description.  It was found thereby
~\cite{Allen02,Gweon03} that the high T ARPES spectra could be
generally well described by the theory lineshapes for the
measurement T and for an \a\ value the same as determined directly
from the angle integrated spectrum.  These results can be seen in
Figs. 7 and 8 of Ref.~\cite{Allen02} and Figs. 4,5 and 6 of
Ref.~\cite{Gweon03}.  Fig.~7 of Ref.~\cite{Allen02} sketches the
bands and shows ARPES data for T=250 K.  Fig.~8 of
Ref.~\cite{Allen02} compares the data to theoretical lineshapes
calculated for T=250 K, for \a=0.9 and for a range of values of
$v_c\over v_s$. The lineshapes include the experimental broadening
in k and $\omega$.  The choice of $v_c\over v_s$=2 gives the best
agreement with the data.  Fig. 6 of Ref.~\cite{Gweon03} shows a
similar comparison of the data to theory for a range of values of
\a~ and one can see that \a=0.9 gives the best agreement for the
rate of falloff of the holon peak intensity as k approaches \kF.
Fig. 7 of Ref.~\cite{Allen02} is interesting for showing explicitly
that k-integration of ARPES data for a Fermi liquid material yields
a Fermi edge whereas one obtains a power law at \EF\ for LiPB.

Fig.~8 of Ref.~\cite{Allen02} (same as Fig.~5 of
Ref.~\cite{Gweon03})  makes an important point concerning the ARPES
lineshapes.  For \a~$>$ 0.5, as is the case for LiPB, the spinon
feature of the lineshape is an edge singularity rather than a peak
singularity.  Including also the broadening due to temperature and
experimental resolutions yields a lineshape which at first glance
does not have two distinct features and so could arouse skepticism
as to the claim of observing spin-charge separation.  Nonetheless
the various panels of the figure make it clear that the two features
are indeed present and are still very visible because the spinon
edge disperses at a rate that differs from that of the holon peak,
controlled by varying $v_c\over v_s$.   Thus these lineshapes are
quite unique, are characteristic of LL behavior, and are much
different from the usual dispersing peaks seen in ARPES.

Two other important findings for the ARPES lineshapes of LiPB are
that the same spectra are obtained \cite{Gweon04} for samples made
using each of the two main crystal growth techniques and that,
within experimental resolutions, the same spectra are obtained when
measured \cite{WANG06PRB} at a photon energy \hv=500\,eV as when
measured with the lower photon energies \hv\ between 20 eV and 30 eV
that were used for all the other ARPES summarized here.  The
significance of the former is that it dispels the possibility of
sample growth method being the origin of an early ARPES report
\cite{Xue99} of Fermi liquid lineshapes for T above T$_{min}$ and of
an 80 meV gap for T below T$_{min}$. This anomalous finding was
disputed \cite{comment, reply} and since then has never been
replicated.  The significance of the latter is that the higher
photon energy spectra are more bulk sensitive and the agreement of
the spectra is consistent with the likelihood based on the crystal
structure that the quasi-1d chains lie two layers below the cleavage
plane and are hence well protected from surface effects, the same
situation that has made ARPES relevant to the bulk properties of
many superconducting cuprate materials.

Scanning tunneling spectroscopy (STS) \cite{Hager05} has made an
important contribution to the case for LL behavior.  This work is
notable for its T-range from 55 K down to 5 K, only slightly above
T$_{SC}$.  The experimental resolution deduced from the STS spectra
is 9 meV, essentially the same as the gap value implied by a mean
field transition at T$_{min}$.  STS also observes a power law
density of states at \EF, although the values of \a~$\approx$~0.6
are smaller than those measured in the high temperature ARPES.  No
change was detected in the T-range of the resistivity upturn,
consistent with the lack of a gap found in optical spectroscopy.
As with ARPES it is likely that the quasi-1d chains being probed in
this surface sensitive spectroscopy are protected from surface
effects by lying well below the cleavage plane.

Photoemission measurements extended to lower temperatures have
continued to support LL physics, but have also revealed interesting
disagreements with TL-model predictions. Lineshapes from angle
integrated ARPES data from 300 K down to 30 K are very well fitted
by TL-lineshapes but only if \a\ is T-dependent \cite{WANG06}. The
T-dependence of \a\ nicely connects the high T value of $\approx$
0.9 to the low T-value of $\approx$ 0.6 found in STS. As with STS
there is no evidence of the resistivity upturn.  However, the fact
that \a\ is T-dependent is clearly outside the one band TL-model. A
microscopic theory offered in Ref.~\cite{WANG06} is that Coulomb
interactions involving the two bands of the two chains per unit cell
(implying four collective modes, two holons, two spinons
\cite{Giamarchi, Wu03}) give rise to a T-dependent renormalization
of \a.

\begin{figure}[t]
\begin{indented}
\item[]\includegraphics[width=0.6\textwidth]{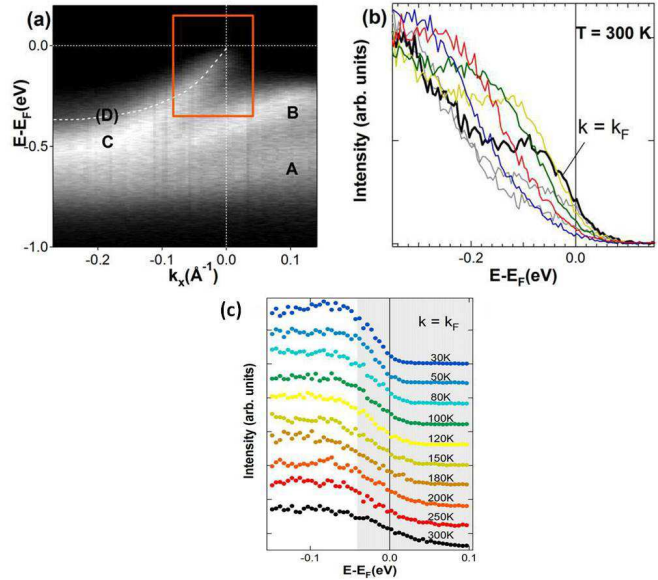}
\caption{ARPES spectra of Li$_{0.9}$Mo$_{6}$O$_{17}$. (a) Intensity map of 300K spectra of T-dependent data set analyzed quantitatively in Ref. \cite {Wang09}.  D is shown as a dashed line because it is strong \cite {JDD99, Gweon01, WANG06} only for different experimental geometries and k-paths. (b) Overplot of spectra in range of box in (a). Different colors represent different $k$ values with increments 3.6 \% of $\Gamma$-Y. (c)  Spectra for $k=k_{\rm F}$ normalized as described in Ref. \cite {Wang09} and in text.} \label{Fig2}
\end{indented}
\end{figure}

T-dependent ARPES lineshapes measured over the T range 300 K to 30 K
showed  QC-scaling \cite{Wang09}.  However the T-dependence observed
was not in agreement with expectations from the TL-model.  As shown in Fig.
1(b-d) of Ref.~\cite{Wang09}, although the lineshapes sharpened
considerably with decreasing T, the sharpening was not as great as
predicted, even when taking account of the measured T-dependence of \a.

We now summarize aspects of the T-dependent ARPES study that are important for the later discussion of Section 5 .  Fig. \ref{Fig2}(a) shows the 300K spectra for wavevector k varying along the $\Gamma$-Y direction of the Brillouin zone.  Bands A and B approach $E_{\rm F}$ no closer than 0.12 eV. Bands C and D merge and disperse to cross $E_{\rm F}$ together.  For the particular k-path shown, the D band is too weak to observe and so its dispersion is sketched as a dashed line based on data \cite {JDD99, Gweon01, WANG06} from other k-paths. Fig. \ref{Fig2}(b) overplots the spectra in the range of the box of Fig. \ref{Fig2}(a) to show the dispersing holon peak and spinon edge approaching $E_{\rm F}$.  The need to measure for multiple T values within the lifetime of the sample necessitated noticeably poorer statistics for these spectra than for the spectra (cited above \cite{WANG06, Allen02, Gweon03}) that were compared in detail to the TL theory lineshapes.  But apart from the poorer statistics the general features of the spectra are the same.

\begin{figure}[t]
\begin{indented}
\item[]\includegraphics[width=0.6\textwidth]{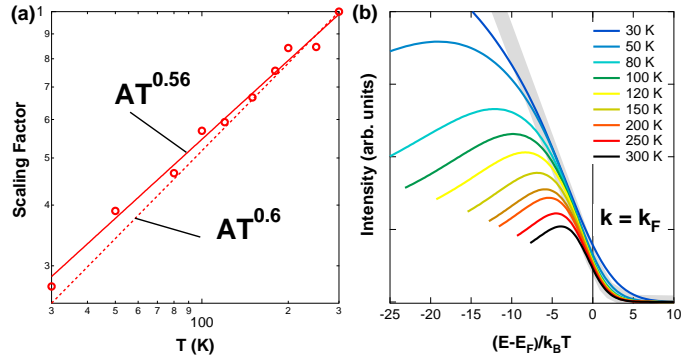}
\caption{(a) Log-log plot of T-dependence of inverse of scaling
factors(circles) needed to obtain Figs. 3(b) and (d) in \cite{Wang09}, showing power law nearly $T^\alpha$. (b) The T-scaling of $k_F$ spectra in the phenomenological description after the intensity is multiplied by a factor of 1/T$^{0.6}$. Direct comparison to data is given in Fig.~\ref{Fig5} as discussed in text. } \label{Fig3}
\end{indented}
\end{figure}

The T-dependent spectra were tested for QC-behavior as follows.  If the general scaling form of the spectral function holds, then  $T^{-\eta} A(k,\omega,T)$ is independent of T if k is chosen for each T so that $k/T$ does not change, i.e. $k =0$ (the $k_{\rm F}$ spectra) or $k=c\,T$ where $c$ is a constant. As described in Ref.~\cite{Wang09} the $k_{\rm F}$ spectra were normalized to one another by matching the leading edges (0.25 to 0.4 eV) of the B band peak, which lies relatively far from $E_{\rm F}$ and has no apparent T dependence.  Fig. \ref{Fig2}(c) shows the normalized spectra vertically offset for clarity.  The spectral intensitives at $E_{\rm F}$ were then matched with a multiplicative T-dependent factor whose inverse has a power law dependence from which $\eta$ can be deduced as shown in Fig.~\ref{Fig3}(a).  The deviations between the data and the power law fit have no systematic pattern and are consistent with the effects of the noise in the spectra (see Fig. \ref{Fig2}(b)) giving uncertainty to the normalization process leading to Fig. \ref{Fig2}(c).   When the $k_{\rm F}$ T-scaled spectra are plotted vs. $(E-E_{\rm F})/k_{\rm B}T$ the spinon edges align.  The scaling behavior of the spectra can be visualized from Fig.~\ref{Fig3}(b) which shows for $k =0$ the result of a phenomenology that is presented in Section 5 and that gives a good description of the data of Ref.~\cite{Wang09}.   We see that the spinon edges scale, but not the holon peaks.  Unscaled k-dependent spectra were normalized and T-scaled by exactly the same factors as already found for the $k_{\rm F}$ spectra and also showed the same general scaling result.  In addition to the lack of scaling of the holon peaks, one sees from Fig.~\ref{Fig3}(a) that $\eta$ has a value that is essentially \a\ rather than the value (\a~- 1) that  would be expected from the TL model \cite{Orgad01}.  The difference between the fitted power law 0.56 and the measured value of \a\ is not deemed significant in view of the uncertainty in the normalization process described above.   On the one hand the observed scaling behavior confirms a generic property of LL physics, but on the other hand the deviations from expectations in the TL model raise difficult questions for 1d theory \cite{viewpoint}.

This section closes by calling attention to another very important
evidence of LL behavior that has recently been reported
\cite{Hussey2011}. A pioneering measurement of the ratio of the
thermal and electrical Hall conductivities in the T range 300 K to
25 K observed a spectacular violation of the Wiedemann-Franz law.
As explained in Ref.~\cite{Hussey2011} the spin-charge separation of
LL physics offers a very natural explanation for this failing
because heat transfer proceeds by both spin and charge degrees of
freedom whereas electric current involves charge alone.  This work
is notable for being clearly bulk sensitive and for demonstrating a
general transport technique capable of distinguishing between Fermi
liquid and LL behaviors.

\section{New Photoemission Data Just Above T$_{SC}$ }

Up to now the only single particle spectroscopy data for
temperatures not far above T$_{SC}$ = 1.9 K has come from the STS
measurement cited above \cite{Hager05} for which the lowest T was 5 K.  Our previous T-dependent photoemssion studies,  angle integrated spectra
shown in Fig. 2 of Ref.~\cite{WANG06}, and ARPES spectra
shown in Fig. 1 of Ref.~\cite{Wang09}, extended from 300K down to only 30K.  In
this section we present and discuss two new sets of photoemission
data taken at temperatures comparable to those of the STS data.  These new data provide a much
higher resolution view of the FS than we have previously published in Fig. 1 of Ref.~\cite{JDD99}
and also allow a more direct comparison to the low T
STS results.  We note that it is very challenging to achieve very low
temperatures with ARPES because of the complexities of sample
holders with multiple mechanical degrees of freedom, and the
difficulties of providing thermal shielding of the sample while also
allowing for photon excitation and electron collection.

The new data were taken with two different ARPES setups where the
lowest temperatures were T=4 K and T=5 K respectively.  One setup is
situated in the laboratory of the Department of Material Physics,
Graduate School of  Engineering Science of the University of Osaka.
Here we obtained angle integrated data with a photon energy of 8.4
eV and a resolution of 5\,meV . The other setup is the MERLIN
Beamline 4.0.3 at the Advanced Light Source (ALS) synchroton. Here
we measured with h$\nu$=30 eV and a resolution of 12\,meV. For both
experiments, single-crystal \LPB\ samples were grown using the
temperature gradient flux method \cite{Greenblatt84}. The
instrumentation in the Osaka laboratory consists of an MB Scientific
MBS T-1 microwave excited rare gas lamp monochromatized by use of
ionic crystals (CaF$_2$ for Kr and sapphire for Xe) \cite{Suga2010}.
Here Xe was used for the lamp which gives a photon energy of
\hv=8.4\,eV.  Other instrumentation consists a low temperature
closed cycle He cryostat manipulator and a Scienta SES2002 electron
energy analyzer driven by an MBS A-1 power supply.  The beamline at
the ALS utilizes a low temperature 6-axis sample manipulator cooled
with an open-cycle He flow cryostat and a Scienta R8000 kinetic
energy analyzer. The MERLIN beamline has a elliptically polarized
undulator which allows one to choose arbitrary polarizations of the
incident light. Due to the longer elastic escape depth of lower
kinetic energy electrons the bulk sensitivity of data from the Osaka
measurements with h$\nu$=8.4 eV is expected to be higher than that
for the synchrotron measurements with h$\nu$=30 eV. The drawback of
that enhanced bulk-sensitivity is a lower cross-section of the
photoemission process, which requires longer acquisition times
compared to the synchrotron in order to achieving sufficiently good
statistics.  In the present case this difficulty was exacerbated by
the small size of the sample relative to the photon spot size of the
laboratory setup and the acquisition time difference was $\approx$24
hours compared to 1 hour.  On the other hand, the temperature for
the laboratory measurement is essentially the same as the lowest T
of the STS measurement while its resolution is even higher and would
be capable of detecting the $\approx$~8 meV mean field gap of a
transition at T$_{min}$.

\begin{figure}[t]
\begin{indented}
\item[]\includegraphics[width=0.8\textwidth]{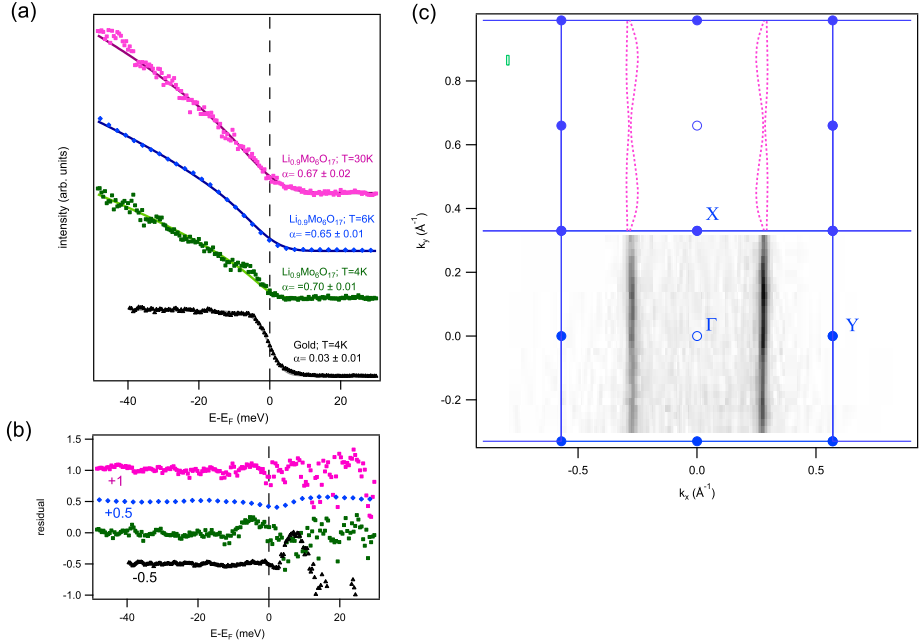}
\caption{(a) Angle integrated photoemission spectra of Lithium
purple bronze for T=4 K, 6K and 30 K. The spectra at T=4K and 30 K
were taken with a resolution of 5 meV with h$\nu$=8.4 eV. For
reference a gold spectrum with the same settings is also shown.  The
angle integrated spectra for T=6 K stems from the same data-set as
the Fermi surface depicted in (c). This data-set was taken with
energy resolution of 12 meV and h$\nu$=30 eV. All the spectra are
generally well fit by the TL model lineshape. The gold spectrum fits
well with \a\ essentially zero, corresponding mathematically to a
Fermi edge. Panel (b) shows the normalized residuals for each fit.
Panel (c) shows the Fermi surface produced by integrating ARPES
spectra $\pm$ 6 meV around the Fermi energy. The green box in the
upper left represents the FWHM of the resolution function in
k-space. The black color means higher intensity. One sees clearly a
very straight Fermi surface represented by the two vertical black
lines. The upper BZ shows the Fermi surface as calculated by
Popovic\cite{POPOVIC06}. Compared to the theoretical FS, there is no
sign of a splitting or warping of the experimental Fermi surface. }
\label{Fig4}
\end{indented}
\end{figure}

We first present the new ARPES results obtained at the ALS.  The sample was
orientated such that the 1-d chains (b-direction) were along the
angular axis of the detector, which is along $\Gamma$-Y in
Fig.~\ref{Fig4} (c) and was cleaved {\em in situ} on the cold
cryostat in a vacuum better than 8$\times$10$^{-11}$ Torr.  The
photon polarization was chosen with the electric field vector
perpendicular to the b-direction. The temperature was maintained at
T=6 K.  A three-dimensional spectroscopic data set I(\kx, \ky, E)
was obtained by rotating the sample around the polar-direction.  The
FS map shown in Fig.~\ref{Fig4} (c) was produced by integrating the
photoemission signal $\pm$6 meV around \EF, essentially the width of
the energy resolution. The k-resolution is 0.01 \AA$^{-1}$ along the
analyzer slit ($\Gamma$-Y) and 0.03 \AA$^{-1}$ perpendicular to the
slit (along $\Gamma$-X), much better than for our previously
published FS map \cite{JDD99}.  Black means high intensity in the
map and we see two vertical black lines which represent the FS.  In
contrast to the theoretical calculations by Popovic et
al{.}\cite{POPOVIC06}, we see at this resolution no splitting of the
FS. Also it appears that the FS is essentially straight at this
resolution.  A Lorentzian peak-fit estimates the value of \kF\ along
the FS to vary by not more than 0.006 \AA$^{-1}$, which is less than
the k-space resolution and less than the variation of the LDA
calculation.

In order to compare with the STS data and for comparison to the
Osaka data presented below,  we angle-integrated the ALS ARPES data.
The result is shown in Fig.~\ref{Fig4} (a). One sees the excellent
statistics of these data.  A line-fit of the spectrum is indicated
by the solid line through the data points.  This fit was made using
the k-integrated spectral weight of a spin-rotational invariant
TL-model \cite{Orgad01} with $v_c\over v_s$=2, consistent with
previous results described above. The theoretical spectrum was
broadened by the experimental energy resolution of 12\,meV.  In
panel (b) of Fig.~\ref{Fig4} one sees the normalized residual of the
fit, i.e. the difference between the fit and the actual spectrum
divided by the actual spectrum.  The residual is essentially smooth.
The  \a~ value of $\approx$ 0.65 is comparable to that obtained at
30 K in our previous photoemission experiments \cite{WANG06, Wang09}
As found with STS \cite{Hager05} there is no significant difference
for T above and below $T_{min}$ at the resolution of this
measurement.

We discuss the Osaka results next.  The samples were oriented by
x-ray diffraction with the chains (b-direction) along the angular
axis of the analyzer. They were cleaved {\em in situ} on the cold
cryostat manipulator in a vacuum better than 2$\times$10$^{-10}$
Torr. As the acceptance angle of the spectrometer is $\pm$ 7 degree
corresponding to $\pm$ 0.11\,\AA$^{-1}$ at \hv=10 eV, and the
Brillouin-zone extends along $\Gamma$-Y to 0.57 \AA$^{-1}$ in the
chain direction, corresponding to $\approx$ 37 degree, we measured
at angles of 0, 14 and 28 degree with the same conditions of time,
photon flux and temperature, and afterwards added the spectra to
obtain angle integration.  After finishing the measurements the
Fermi-energy was determined by evaporating gold onto the sample. The
spectra of LiPB for T=4 K and 30 K are shown in Fig.~\ref{Fig4} (a)
together with the corresponding gold spectrum at T=4\,K.  One sees
the considerably poorer statistics of the LiPB Osaka data relative
to that of the ALS data or the gold spectrum.  Nonetheless the sharp
contrast between the reduced weight near \EF\ in the LiPB spectra
and the Fermi edge of the gold spectrum can clearly be seen. Solid
lines through the data indicate line-fits of the spectra using the
same TL-model as for the ALS k-integrated spectrum.  The theoretical
spectrum was broadened by the experimental energy resolution of
5\,meV which was determined from the gold spectrum. The \a\ value of
the fit is written beside each spectrum together with the
standard-deviation resulting from the $\chi^2$-fit.   For the gold
spectrum, we expected a Fermi edge and therefore \a=0. This small
deviation from zero and also the peak above \EF\ in the residual
shown in Fig.~\ref{Fig4}(b) might be seen as indicating some
systematic error.

In panel (b) of Fig.~\ref{Fig4} one sees the normalized residuals of
the fits.  The numbers next to the curves indicate how much the
curves were shifted in the vertical direction for improved clarity
of presentation.  Comparison of the residuals again shows clearly
the poorer statistics of the Osaka LiPB data.  Both the T=4 K and
the T=30K spectra show a deviation from the TL-lineshape at around 5
meV binding energy.   This deviation is larger for T=4 K than for
T=30 K.  Since the resolution of 5 meV is better here than in the
STS experiment it is logically possible that such a feature could
have escaped detection in the STS spectrum.  But a skeptic might
point out that because it appears also in the 30K spectrum it does
not correlate directly with the T-dependent transport properties.  A
skeptic might also claim a hint of this feature even in the gold
spectrum residual.   Thus the question of whether this deviation is
real or is due a small systematic error, i.e. not perfectly linear
behavior of the detector or a contribution from the sample holder
because the photon spot is large, or is simply the result of
insufficient acquisition time to obtain better statistics, cannot be
absolutely answered.  We present the spectra as a further
confirmation of the overall large energy scale power law behavior
and regard them as ambiguous on the possibility of structure at the
5 meV energy scale of the resolution.  We take it as significant
that all the new LiPB data sets presented here give essentially the
same \a~ values $\approx$ 0.65 to 0.7. This is very much consistent
with the range of values found in STS and perhaps consistent with a
slight low T upturn of \a\ found previously, as shown in
Fig.~\ref{Fig3}(a) of Ref.~\cite{WANG06}.

\section{Phenomenological Description of T-dependent ARPES Data}

The findings of Ref.~\cite{Wang09} show that \LPB\ is a QC system,
but with important differences relative to expectations from the
one-band TL-model, specifically in the exponent of the temperature
prefactor and in the lack of the full sharpening predicted for
decreasing T, which is part of the non-scaling of the holon peaks.

These differences can be partially described by a phenomenological
momentum broadening of the TL spectral function, but the required
broadening greatly exceeds the experimental momentum resolution, as
discussed below.  The fact that the experimental scaling prefactor
is $T^\alpha$ rather than $T^{(\alpha-1)}$ is a basic motivation to
try an integration of the theoretical TL spectral function because
that will draw a factor of T outside the integral.  Convolving the
theoretical spectral function (for which $\eta$ = (\a~- 1)) with a
momentum window function $R(p/p_0)$, where $p_0$ is a T-independent
width, and using the change of variables $\tilde{p} = \frac{(v)(p)}{
T}$ (recall that $v$ is a constant with units of velocity), gives
\be A_{\rm test}(k,\omega, T) = T^{\alpha} \int_{-\infty}^{\infty}
\tilde{R}\left(\frac{\tilde{p}}{\frac{v p_0}{T}}\right)
\tilde{A}(\tilde{k}-\tilde{p},\frac{\omega}{T}) d\tilde{p}
\label{new_scaling} \ee A constant factor of 1/$v$ has been drawn
into a redefined $\tilde{A}$.
 $A_{\rm test}(k,\omega,T)$ has the
$T^{\alpha}$ prefactor observed experimentally but the rest of the
expression is no longer a universal function of $\frac{\omega}{T}$
because T also enters in the form $\tilde{p_0} = \frac{v p_0}{T}$
and $A_{\rm test} = T^{\alpha} f(\frac {v p_0}{T},\frac{v
k}{T},\frac{\omega}{T})$.  It can be noted in passing that this
exercise also shows the way in which a fixed experimental resolution
has an increasingly large effect on a QC spectrum as temperature
decreases.

\begin{figure}[b]
\begin{indented}
\item[]\includegraphics[width=0.6\textwidth]{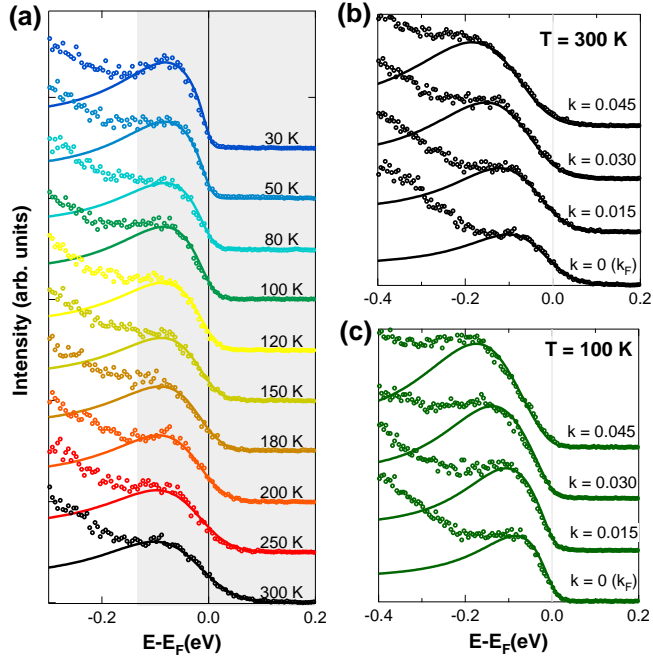}
\caption{Fitting the ARPES data of Ref.~\cite{Wang09} to $A_{test}$. (a) Fitting for $k$ =\kF\ at various temperatures. (b) and (c) Fitting to both $k$ = \kF\ and $k$ away from \kF\ at 300 K and 100 K respectively. Parameters are: $v_c\over v_s$=2 and \a = 0.6, $v_{s} = 1.9$ eV\AA\ and $p_0$= 0.065 \AA$^{-1}$, same for all panels. Unit of $k$ is \A-1.}
\label{Fig5}
\end{indented}
\end{figure}

Choosing $R$ to be a simple normalized Gaussian with p$_{0}$ = 0.065
\AA$^{-1}$ and choosing the low temperature value of \a = $0.6$,
consistent with the experimental prefactor, yields a reasonable
description of the experimental data below the binding energy for which
the intensity from band B becomes important. Fig.~ \ref{Fig3}(b)
shows the general scaling behavior of $A_{\rm test}$, including the
experimental energy resolution.  After the intensity is multiplied
by 1/T$^{\alpha}$ and the energy axis is scaled by $k_B$T, the
theory curves at \kF\ have their edges fall on each other while the
peak parts do not, exactly what is observed in experiment. The
slight deviation from the scaled edge at low temperature above \EF\
is seen in the data of Ref.~\cite{Wang09}, and as shown there, is
due to the experimental resolutions.

Fig.~\ref{Fig5}(a) compares the \kF\ data of Fig.~\ref{Fig2}(d)
for 10  values of T with the phenomenological $A_{\rm test}$ and Figs.
5(b,c) make the comparison for four values of k, T-scaled away from
\kF\, at two temperatures.   In these figures, the amplitudes of the
experimental curves have been adjusted to remove the slight (maximum
15\%) prefactor differences from $T^{\alpha}$ seen in
Fig.~\ref{Fig3}(a) and ascribed in the Section 3 discussion of that figure to uncertainty in the normalization process.  By eye there is generally good agreement in
these comparisons until reaching the binding energy at which the
higher lying bands begin to contribute intensity in the experimental
spectra.  From the spectra of Fig.~\ref{Fig2}(a) and taking account of the width of band B this energy is 0.1 eV to 0.12 eV.  For k-values away from \kF\ the binding energy of band B increases and so the deviation also moves to higher binding energies, nearly 0.2 eV.  In spite of this agreement,  it is important to point out that this phenomenology
does not produce the T-dependence of \a~ that is measured from T-dependent k-integrated spectra.   In fact
it works only for k-values rather near to \kF~ whereas a much larger
range of k contributes to the k-integrated spectrum that determines
\a.  Thus the phenomenology is not a substitute for a satisfactory
microscopic model.

What is the meaning of this phenomenology?  The simplest possibility
is of course that the momentum is indeed broadened. One could argue
that the QC scaling predicted in the TL model is being observed
except for being modified by the experimental resolution. However,
the value of p$_{0}$ is 5 times larger than the instrumental $k$
resolution, computed from the instrumental angle resolution as 0.013
\AA$^{-1}$.  Such a large extra broadening would have to be ascribed
to sample surface quality. For example our relatively large spot
size might be illuminating regions with slightly different
orientations on the sample surface. The size of the discrepancy
makes this interpretation more difficult to defend, but nonetheless
the phenomenology shows that the issue of finite momentum resolution
is very important.  ARPES scaling data might evolve with progress in
sample growth and surface preparation techniques.

Another possibility is that the phenomenology tries to catch some
of the intrinsic physics.  For example it is clear from the
T-dependence of \a~ that the one band TL model is missing
interactions of importance and we know from band theory and the
ARPES data that there are two bands crossing \EF, associated with
the two chains per unit cell.   Perhaps the phenomenology can be a
guide for future theory.  It would be very desirable to test the
data against the spectral function for a two band TL model if it
were available.

\section{Robust 1d Physics}

The combination of data from ARPES and STS along with the
Wiedeman-Franz law violation cited above firmly establish generic
signatures of LL physics in LiPB for T at least down to T$_{min}$.
How does the system achieve the 3d character that must obtain in the
SC state? At this point it is not possible to make a unified
interpretation of all the relevant data.  One can only speculate in
a general way, guided by current 1d theory.  Recent discussions
\cite{NeumeierThermal07, Neumeier08, Hussey09, Chen2010,
Neumeier2011} focus on the idea that the crossover to 3d occurs
before T$_{SC}$ is reached, either that the resistivity rise is due
to CDW formation \cite{NeumeierThermal07, Hussey09}, which is
certainly of a 3d nature, or that a CDW does not occur
\cite{Neumeier08, Chen2010, Neumeier2011} and that crossover is more
subtle in the resistivity data \cite{Chen2010} and in the mechanism
whereby it occurs \cite{Neumeier08, Neumeier2011}.  Here we
speculate that 1d physics may be more robust, with crossover to 3d
occurring only with the transition to SC, i.e. that above T$_{SC}$
there is non Fermi liquid behavior rooted somehow in 1d with no
single particle gap or density wave.  In this connection one can
take note of recent arguments \cite{HusseyHc2} for unconventional
SC, possibly spin-triplet, based on the finding that the critical
field is much larger than the Pauli limit.

Because the 1d physics of LiPB appears to lie outside standard 1d theory \cite{viewpoint} we can only try to identify generic possibilities that may be robust for this line of thinking.  Looking first to single particle spectroscopy, the power laws observed for temperatures only slightly above T$_{SC}$ in the STS and ARPES data with 12 meV resolution are an initial temptation to consider this hypothesis.  As discussed in Section 4, the 4 K angle integrated spectra are ambiguious on the possibility of spectral structure on a lower energy scale of the 5 meV resolution, but we note that even this energy scale corresponds to a temperature of 58 K, much higher than the T of the measurement or the T$_{min}$ of the resisitivity upturn.  The energy scale can perhaps be pushed lower by the 6 K optical spectroscopy \cite{Degiorgi88} performed down to 1 meV, corresponding to 11.6 K.  This work found no single particle gap and no change associated with the resistivity upturn but did employ a Drude description of the infrared data.  A more recent study \cite{Choi04}, albeit a measurement down to T=10 K and with a lower energy limit of 6 meV, also found no gap but in addition showed that the infrared behavior is actually non-Drude.    Taken together these two studies suggest no single particle gap and non-Fermi liquid behavior down to an energy scale of 1 meV, but definitive low temperature spectroscopic measurements probing to energy scales well below 1 meV are clearly needed.

The resistivity upturn, with the possibility of a gap less than 1 meV, as obtained in Ref.~\cite{Hussey09} and shown in Fig.~\ref{Fig1}(a), is the primary transport evidence suggesting crossover due to a small single particle gap.  However the highly novel gap function and the T-independent constant dc magnetic susceptibility measured down to 2 K are cause to consider an alternative description of the upturn.

Power laws are characteristic features of the QC nature of LL physics.  Indeed a power law is expected for the resistivity of a LL system.  Fig.~\ref{Fig1}(a) shows that the data of Ref.~\cite{Hussey09} can be described by two power laws, one with positive exponent for the metallic part \cite{linear} and one with negative exponent for the upturn.  Fig.~\ref{Fig1}(b) shows the T-dependent gap that results from recasting this fit as an Arrhenius law.  The gap so obtained is very similar to that obtained in Ref.~\cite{Hussey09} and the difference is very likely due to the difficulty of accurately digitizing the upturn in the resistivity data.  In this view the unusual gap function is the result of forcing an Arrhenius description onto a power law.  In particular one notes that at low T the gap must roll over and then actually decrease in order to map the very fast rising exponential onto the slower rising power law.

Such a two-power-law description of LiPB resistivity was first put forth in Ref.~\cite{Neumeier08}, where the two power laws were ascribed to two independent LL's, one for each of the two bands crossing \EF\, with each having different values of \a\ and quite different roles in the crossover physics.  This interpretation neglects the Coulomb interaction that couples the chains to give symmetric and antisymmetric holon and spinon modes, as discussed in Ref.~\cite{Giamarchi} and Ref.~\cite{Wu03} and applied to LiPB in Ref.~\cite{WANG06}.  An alternate LL interpretation for the power law upturn is the old proposal of Anderson localization due to disorder, but put in the context of LL physics as a low T crossover of the sign of the resistivity power law exponent due to disorder \cite{Giamarchi88}.  A great concern of early discussions \cite{MatsudaXiNoCDWLoc86, MatsudaMixedCrys86, SatoLoc87, EkinoTunneling87} was that disorder and localization could be incompatible with SC.  But, as suggested in Ref.~\cite{EkinoTunneling87}, the increased sensitivity to disorder in 1d could make it possible for the energy scale of the disorder to be much less than T$_{SC}$.

As seems true of all hypotheses concerning the resistivity upturn, this one also leaves some aspects of the data unexplained, why the same T$_{min}$ would be found for the resistivity along all three axes, why hydrostatic pressure suppresses the upturn (and also enhances the SC) \cite{SchlenkerPressure89}, and why the upturn can be suppressed  \cite{Hussey09} by a magnetic field.  For these aspects of the data one sees the attraction of postulating a gap that can be affected by pressure or a magnetic field.   Whatever is the proper understanding of the upturn, a case can be made that by virtue of a power law dependence on T it signifies a continuation of 1d behavior rather than a loss.  The highly directional nature of the magnetic field suppression of the upturn \cite{Hussey09} emphasizes the importance of the 1d character in this temperature range and in this connection one notes also the remarkable recent observation \cite{Hussey09} that for a non-superconducting crystal a sufficiently large magnetic field applied specifically along the 1d axis appeared to restore the SC with a transition temperature considerably higher than 1{.}9\,K.

To return to the theme of the opening paragraph of the paper, the greatest issue to be confronted in any line of thinking that emphasizes 1d physics down to T$_{SC}$ for LiPB is the role and the energy scale of $t_{\perp}$. The data of Fig.~\ref{Fig4}(c) show that the FS is unsplit and straight to a greater extent than predicted in current LDA calculations.  If one accepts the current lack of evidence for a CDW, and that is the stance of this paper, then there is a need to understand why the good nesting of the FS does not result in a CDW having all the standard properties that are readily detected for CDWs in related materials. Whatever is the mechanism it would apply down to T$_{SC}$.  Within present theory the situation is very difficult but perhaps not quite impossible.  We now discuss three possibilities.

In a one band LL model it is well known that the easy route to avoiding crossover due to $t_{\perp}$ is a gap in the spinon mode.  The gap energy must be overcome for single particle hopping between the chains, which renders $t_{\perp}$ to be irrelevant in the renormalization group sense \cite{Giamarchi}. But pair hopping that is second order in $t_{\perp}$ remains relevant and can lead to SC \cite{Giamarchi}.  If the energy scale of the spin gap is greater than T$_{SC}$ there can be a transition directly from the spin-gapped LL (a Luther-Emery liquid) to a SC \cite{Carlson2000}.  Such a gap in LiPB would seem to be precluded by the T-independent dc magnetic susceptibility which has been measured down to 2 K \cite{Choi04}.  However, the two chain model with four modes, as mentioned in the preceding paragraph, has both a symmetric and an antisymmetric spin mode.  The symmetric mode that is probed in the dc susceptibility can be ungapped for some parameter ranges of this model \cite{Wu03}.  One could think of the possibility of a gap in the antisymmetric spin mode that has not been probed in any transport measurement to date and is small enough to have escaped spectroscopic detection but is nonetheless larger than T$_{SC}$.  Such a gap would also be a barrier against single particle hopping between the chains and so might render $t_{\perp}$ to be irrelevant.

A second possibility is the so-called sliding Luttinger liquid (SLL)
\cite{SLL1, SLL2, SLL3}.  This is a lattice model for an array of
coupled chains and specifically includes a mechanism for suppressing
a purely electronic CDW along the chains. The physical idea
\cite{SLL2, SLL3} is that Coulomb interactions between the chains
lead to transverse incommensurate CDW fluctuations that modulate the
charge density on the chains, and therefore the Fermi wavevector and
therefore the longitudinal CDW q-vector, which frustrates locking of
the longitudinal CDW.  In this special regime the T=0 ground state
retains the LL properties of a single chain and is stable against
$t_{\perp}$, against CDW fluctuations and SC fluctuations.  A spin
gap is not required but of course aids the SLL stability.  But in
any case the parameters of the model must be tuned to bring the
system to be near such a transverse CDW instability, raising doubt
as to whether the model could apply to any real material.  For other
parameter values the stable ground state is a CDW, a novel SC, or a
Femi liquid.  In addition there are difficulties specific to its
application to LiPB.  The model does not address the possibility of
a thermal phase transition from the SLL state to a SC.  Also, as T
goes to zero for the SLL state, the resistivity is predicted to
become infinite transverse to the chains and to become zero parallel
to the chains.  For the transverse direction the resistivity upturn
is then consistent with this prediction but for the parallel
direction one must invoke some other mechanism such as disorder. But
one must then explain why T$_{min}$ is the same for all directions
and one must take account of how the transverse interactions may
change the effect of disorder \cite{SLL3}. Nonetheless the concept
that the chain CDW can be suppressed by frustration arising from
competition between parallel and transverse CDW fluctuations may be
more robust than the details of the model calculations that have
been done and so the SLL remains interesting for LiPB.

In the absence of a spin gap or a scenario like the SLL, the only
possibility for the line of thinking in this section would seem to
be that the energy scale of $t_{\perp}$ is less than T$_{SC}$.  In
that case there is presently no conclusive theory for or against
direct crossover of LL behavior to SC.  The question entails the
interplay of the pair tunneling terms needed for SC and the extent
to which Coulomb interactions might suppress or enhance the pair
tunneling.   Could the energy scale of $t_{\perp}$ be this small in
LiPB?  Current LDA calculations \cite{POPOVIC06, Satpathy} suggest
values of $t_{\perp}$ $\approx$ 30 meV that would preclude this
possibility.  However 1d fluctuations on the chains are known to
produce a suppression of $t_{\perp}$ \cite{Bois,
Giamarchi,t-perpeff} to an effective value
$t_{\perp}$$(t_{\perp}$/t)$^{\alpha/(1-\alpha)}$, where t is the
interchain hopping.  For t $\approx$ 800 meV suggested by LDA bands
for LiPB and the measured low T value of \a~$\approx$ 0.6, one
obtains   the very small effective hopping value of 0.22\,meV,
essentially the same as T$_{SC}$.  Although this estimate is
terribly sensitive to the value of \a, which is not known with such
great precision, nonetheless it shows that the effective hopping
could be very greatly suppressed, consistent with the FS data in Fig.~\ref{Fig4}(c).

To pursue this idea our current ARPES research aims at quantifying the
single particle inter-chain hopping and setting a more precise bound on the
extent of warping and splitting of the FS through yet higher resolution
measurements just above T$_{SC}$.  New LDA calculations
using the NMTO method \cite{NMTO} and downfolding can be used to
characterize the $t_{\perp}$ hopping in greater detail \cite{NMTOCollab}. It
is anticipated that the theory $t_{\perp}$'s can then be adjusted in
a realistic way to describe the ARPES data and that the adjustments
required can be compared to expectations from 1d theory
\cite{1dtheorycollab}.  Thereby it is hoped to obtain a more precise picture of the electronic structure that is
giving rise to the remarkably robust 1d properties and perhaps
unconventional SC of LiPB.

\ack This work was supported at UM by the U.S. National Science
Foundation (NSF) (DMR-03-02825 and DMR-07-04480), at Clemson by the
SC EPSCOR/Clemson University Cost Share and the U.S. Department of
Energy (DOE) Implementation Program, and in Osaka by a Grant-in-Aid
for 21st century COE (G18), Global COE (G10), Innovative Areas
"Heavy Electrons" (20102003), and Scientific Research (18104007,
18684015, 21740229, and 21340101) from MEXT and JSPS, Japan. The
Advanced Light Source is supported by the Director, Office of
Science, Office of Basic Energy Sciences, of the U.S. Department of
Energy under Contract No. DE-AC02-05CH11231. We gratefully
acknowledge discussions with J. V. Alvarez, P. Chudzinski, T.
Giamarchi, S. Kivelson and K. Sun.


\begin{thebibliography}{99}
\bibitem{same} The same is true for the hypothesis of the present paper.
\end{thebibliography}
\end{document}